# Numerical study of natural convection states in a horizontal concentric cylindrical annulus using SPH method

Xiufeng Yang, Song-Charng Kong[*]

Department of Mechanical Engineering, Iowa State University, Ames, IA 50011, USA

[*] Corresponding author: kong@iastate.edu

**Abstract**

Natural convection is of great importance in many engineering applications. This paper presents a smoothed particle hydrodynamics (SPH) method for natural convection. The conservation equations of mass, momentum and energy of fluid are discretized into SPH equations. The body force due to the change of density in a temperature field is considered by the Boussinesq approximation. The numerical method is validated by comparing numerical results with experimental results from literature. The numerical and experimental results reach a good agreement. Then the SPH method is applied to study the natural convection in a horizontal concentric cylindrical annulus with Rayleigh number in the range of $10^2$ to $10^7$ and Prandtl number in the range of 0.01 to 10. In general, the flow is stable at low Rayleigh number but unstable at high Rayleigh number. The transition Rayleigh number from stable to unstable states is lower in the low Prandtl number cases than in the high Prandtl number cases. Four different convection states are identified in numerical simulations, namely, stable state with 1 plume (SP1), unstable state with 1 plume (UP1), stable state with n (n > 1) plumes (SPN), and unstable state with n (n > 1) plumes (UPN). The SP1 and UP1 states are observed for all Prandtl numbers, while the SPN and UPN states are only observed at Pr = 0.1 and 0.01.

## 1. Introduction

Natural convection heat transfer in an enclosed space is an important phenomenon in many thermal engineering applications, such as heat exchanges, solar collect-receivers, and heating or cooling systems. Natural convection is also a fundamental problem in fluid mechanics. As a typical example, the natural convection in a horizontal cylindrical annulus has been studied experimentally and numerically for several decades [1, 2]. Early studies on natural convection between horizontal concentric cylinders experimentally measured the heat transfer coefficients [3-6] and proposed several corresponding correlations [7, 8]. Kuehn and Goldstein [2] experimentally studied the influence of eccentricity on natural convection heat transfer in horizontal cylindrical annuli. Their results showed that the change of eccentricity changed the local heat transfer coefficient significantly while the overall coefficients changed less than 10 percent. Kuehn and Goldstein [1] also studied temperature distributions and local heat transfer coefficients for natural convection with in a horizontal annulus experimentally and numerically. They applied a finite difference method to solve the governing equations. The numerical solutions not only agreed with their experimental results but also extended their investigation to lower Rayleigh numbers.

With the development of computational fluid dynamics methods in recent years, the numerical simulation is used more and more to study natural convection between horizontal annuli. The first numerical study was conducted by Crawford and Lemlich [9], who rewrote the differential equations as finite difference equations to solve the problem. The finite difference method was also applied by Kuehn and Goldstein [1, 10] and Cho et al. [11] natural convection heat transfer in horizontal cylindrical annuli. Zhao and Zhang [12] proposed an unstructured-grid upwind finite-volume method and validate their method by simulating natural convection flows in eccentric annuli. Shu et al. [13] studied natural convection heat transfer in a horizontal eccentric annulus between a square outer cylinder and a circular inner cylinder using differential quadrature method. Abu-Nada [14] investigated the heat transfer enhancement of nanofluid in horizontal annuli using finite volume method. Ashorynejad et al. [15] studied the effect of static radial magnetic field on natural convection in a horizontal cylindrical annulus filled with nanofluid using lattice Boltzmann method. Liang et al. [16] validated their moving particle semi-implicit method by modeling natural convection heat transfer between concentric cylinders.

Natural convection heat transfer in a horizontal cylindrical annulus can be characterized by two dimensionless parameters: the Rayleigh number Ra and Prandtl number Pr, which are defined as

$$\mathrm{Ra} = \frac{g\beta L^3 \Delta T}{\nu \alpha}, \quad \mathrm{Pr} = \frac{\nu}{\alpha} \tag{1}$$

where $g$ is the modulus of gravitational acceleration, $\beta$ is the thermal expansion coefficient of the fluid, $L$ is the gap between the cylinders, $\Delta T$ is the temperature difference between the cylinders, $\nu$ is the kinematic viscosity of the fluid, and $\alpha$ is the thermal diffusivity of the fluid. The Prandtl number is the ratio of viscous diffusivity to thermal diffusivity. The Rayleigh number can be thought of as the ratio of the gravitational potential energy to the energy due to viscous dissipation and thermal diffusion [17]. According to Kuehn and Goldstein [1], the flow in a horizontal cylindrical annulus is steady over the range of Rayleigh number from $10^2$ to $10^5$. Kuehn and Goldstein [2] experimentally studied the flow patterns at Rayleigh numbers from $2.2 \times 10^2$ to $7.7 \times 10^7$. They found that the plume above the inner cylinder began to oscillate when Rayleigh number is near $2 \times 10^5$ and the entire plume was turbulent at $\mathrm{Ra} = 2 \times 10^6$. Labonia and Guj [18] conducted an experimental study of transition from steady laminar to chaotic flow in a horizontal concentric annulus in a range of Rayleigh number from $0.9 \times 10^5$ to $3.37 \times 10^5$ and differentiated several stages in the transition from steady state to turbulence. However, these studies on flow patterns have been limited to the Prandtl number near 0.7.

The primary objective of the present study is to extend our knowledge of flow patterns in a horizontal concentric annulus in a range of Rayleigh number from $10^2$ to $10^7$ and Prandtl number from 0.01 to 10. A secondary objective is to present an alternative smoothed particle hydrodynamics (SPH) method for modeling natural convection flows. The SPH method is a Lagrangian meshfree particle method, which was first proposed to model astrophysical problems [19, 20]. Later, it was used to simulate fluid problems [21-23] and fluid-structure interactions [24-26] without considering heat transfer. In recent years, it was extended to fluid problems considering heat transfer. Cleary [27] presented an SPH method with Boussinesq approximation for simulating convective heat transfer. Szewc et al. [28] developed a variant of SPH method for natural convection flow, which did not apply the classical Boussinesq approximation. Yang and Kong [29] proposed an SPH method for evaporating flows in which heat transfer is of great importance.

The following of this paper is organized as follows. Section 2 gives the numerical method, including the governing equations and numerical formulas. Then the numerical method is validated by comparing with experimental results in Section 3. With different Prandtl and Rayleigh numbers, several sets of numerical simulations are conducted. The results are shown in Section 4 with some discussions. The conclusions are given in section 5.

## 2. Methodology

For viscous flow with natural convection heat transfer, the Lagrangian form of the conservation equations of mass, momentum and energy are used:

$$\frac{d\rho}{dt} = -\rho \nabla \cdot \boldsymbol{u} \tag{2}$$

$$\frac{d\boldsymbol{u}}{dt} = -\frac{1}{\rho}\nabla p + \frac{\mu}{\rho}\nabla^2 \boldsymbol{u} + \boldsymbol{F}^B \tag{3}$$

$$\frac{dT}{dt} = \frac{1}{\rho C_\mathrm{p}} \nabla \cdot (\kappa \nabla T) \tag{4}$$

where $\rho$ is the fluid density, $\boldsymbol{u}$ is the fluid velocity, $p$ is the fluid pressure, and $\mu$ is the dynamic viscosity of the fluid, $T$ is the fluid temperature, $C_\mathrm{p}$ is the specific heat at constant pressure, and $\kappa$ is the thermal conductivity of the fluid. $\boldsymbol{F}^B$ denotes the body force acting on the fluid. According to the Boussinesq approximation, the body force due to the change of density in a temperature field can be calculated as

$$\boldsymbol{F}^B = -\boldsymbol{g}\beta(T - T_r) \tag{5}$$

where $\boldsymbol{g}$ is the gravitational acceleration, $\beta$ is the thermal coefficient of volumetric expansion of the fluid, and $T_r$ is the reference temperature.

The SPH method is a Lagrangian particle method for solving differential equations. This is why the governing equations is written in Lagrangian form. In SPH, the interpolated value of a function $f$ at position $r_a$ is

$$f(\mathbf{r}_a) = \sum_b f(\mathbf{r}_b) \frac{m_b}{\rho_b} W(\mathbf{r}_a - \mathbf{r}_b, h) \tag{6}$$

where the subscripts $a$ and $b$ denote particles, $m$ is the mass of a particle, $W$ is a kernel function, $h$ is a smoothing length used to control the width of the kernel. The summation is taken over all particles. However, due to the width of the kernel function, the summation is only over near neighbors.

To avoid the so-called tensile instability in SPH [30] that may occur in fluid simulations, the following hyperbolic-shaped kernel function in two-dimensional space is used [31, 32]

$$W(s, h) = \frac{1}{3\pi h^2} \begin{cases} s^3 - 6s + 6, & 0 \leq s < 1 \\ (2-s)^3, & 1 \leq s < 2 \\ 0, & 2 \leq s \end{cases} \tag{7}$$

where $s = r/h$.

The gradient of the function $f$ can be obtained by

$$\nabla f(\mathbf{r}_a) = \sum_b f(\mathbf{r}_b) \frac{m_b}{\rho_b} \nabla_a W(\mathbf{r}_a - \mathbf{r}_b, h) \tag{8}$$

This is the basic form of gradient. There are several other variants which can be found in Ref. [33, 34].

Using the particle summations formulas of a function and its derivatives, the governing equations (2)-(4) can be discretized into the following form

$$\frac{d\rho_a}{dt} = \sum_b m_b (\mathbf{u}_a - \mathbf{u}_b) \cdot \nabla_a W_{ab} \tag{9}$$

$$\frac{d\mathbf{u}_a}{dt} = -\sum_b m_b \left( \frac{p_a}{\rho_a^2} + \frac{p_b}{\rho_b^2} \right) \nabla_a W_{ab} + \sum_b \frac{m_b (\mu_a + \mu_b)(\mathbf{r}_a - \mathbf{r}_b) \cdot \nabla_a W_{ab}}{\rho_a \rho_b (r_{ab}^2 + \eta)} (\mathbf{u}_a - \mathbf{u}_b) - \mathbf{F}_a^B \tag{10}$$

$$\frac{dT_a}{dt} = \frac{1}{C_p} \sum_b \frac{m_b (\kappa_a + \kappa_b)(\mathbf{r}_a - \mathbf{r}_b) \cdot \nabla_a W_{ab}}{\rho_a \rho_b (r_{ab}^2 + \eta)} (T_a - T_b) \tag{11}$$

where the subscripts $a$ and $b$ denote SPH particles, $m$ is the mass of a particle. $W_{ab} \equiv W(\mathbf{r}_a - \mathbf{r}_b, h)$, here $h$ is a smoothing length used to control the width of the kernel. $\nabla_a W_{ab}$

denotes the gradient of $W$ taken respect to particle $a$. The term $\eta = 0.01h^2$ is added to prevent the singularity when two particles are too close to each other [35].

The above equations are not closed, thus the following equation of state is used to obtain pressure

$$p = c^2(\rho - \rho_r) \qquad (12)$$

where $c$ is a numerical speed of sound and $\rho_r$ is a reference density.

For SPH simulation, the density and pressure fields may undergo large fluctuations numerically. In order to reduce the fluctuation, the Shephard filtering [36] is applied to reinitialize the density field every 20 time steps.

$$\tilde{\rho}_a = \frac{\sum_b m_b W_{ab}}{\sum_b V_b W_{ab}} \qquad (13)$$

## 3. Validation

A schematic diagram of the cross section of a horizontal annulus is shown in Fig. 1. The diameter of the inner cylinder is $D_i$ and the spacing between the two concentric cylinders is $L$. The temperatures of the inner and outer cylinders are $T_i$ and $T_o$ ($T_i > T_o$), respectively. The annulus is filled with fluid. Three non-dimensional parameters, namely, $L/D_i$, Pr and Ra, are set to be the same as that in experiments [1, 2], thus the numerical results can be compared with the experimental results. For all the cases in this paper, $L/D_i = 0.8$.

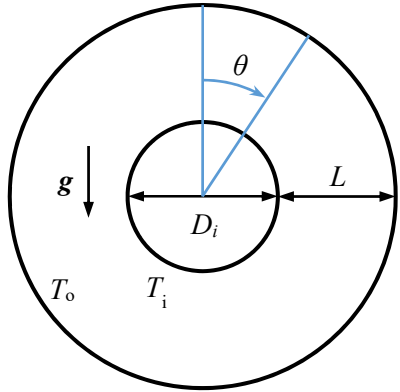

Fig. 1. Sketch of the cross section of a horizontal annulus. $\theta$ is defined as the angle from the top vertical position in clockwise direction.

In the annulus, the hot fluid tends to move up while the cold fluid tends to move down because of the body force due to the change of fluid density in a temperature field. As shown in Fig. 2, the temperature of the fluid near the hot inner boundary is higher than that near the cold outer boundary, thus the hot fluid moves up along the inner boundary and forms a plume at the top center part of the annulus, while the cold fluid moves down along the outer boundary to the bottom. It can be seen in Fig. 2 that the SPH simulation agrees well with the experiment [1] and CVFEM simulation [37]. A quantitative comparison of temperature profiles between SPH simulation and experiment [1] is shown in Fig. 3. They reach a good agreement in different directions from top to bottom.

The SPH simulation is valid at not only low but also high Rayleigh numbers. Fig. 4 compares the interferograms from experiments [2] and the isotherms from SPH simulations at three different Rayleigh numbers higher than that of the previous case. It can be seen in Fig. 4 that the SPH results are in agreement with the experimental results. With the increase of the Rayleigh number, the natural convection of the fluid becomes stronger, and the plume of the hot fluid becomes narrower. The plume even becomes unstable when the Rayleigh number is $1.7 \times 10^7$.

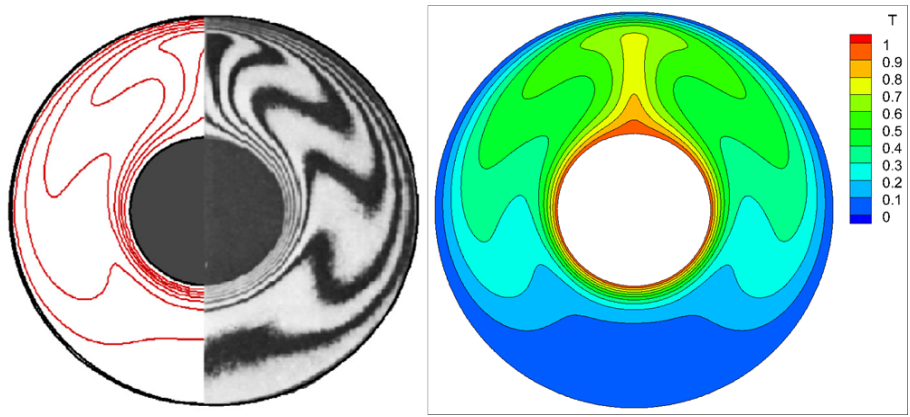

Fig. 2. Comparison of isotherms from CVFEM simulation (left part of the left figure) [37], interferograms from experiment (right part of the left figure) [1] and isotherms from SPH simulation (right figure) (Pr = 0.706, Ra = $4.7 \times 10^4$).

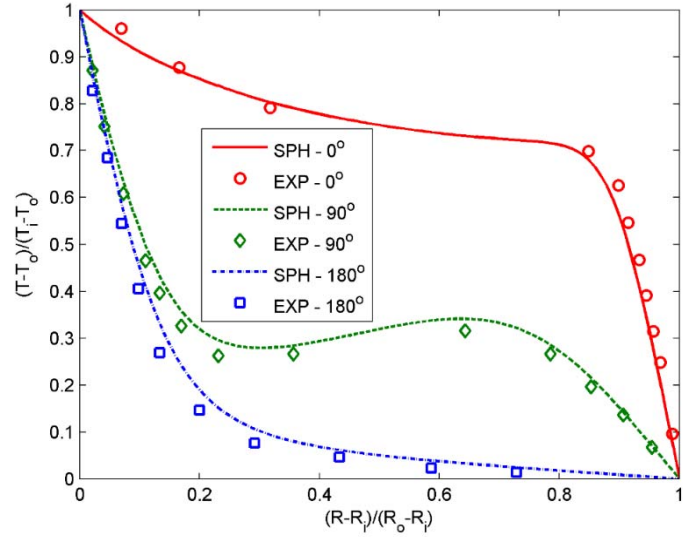

Fig. 3. Comparison of dimensionless radial temperature profiles between experiment [1] and SPH simulation. The degrees in the legend are the values of $\theta$ defined in Fig. 1.

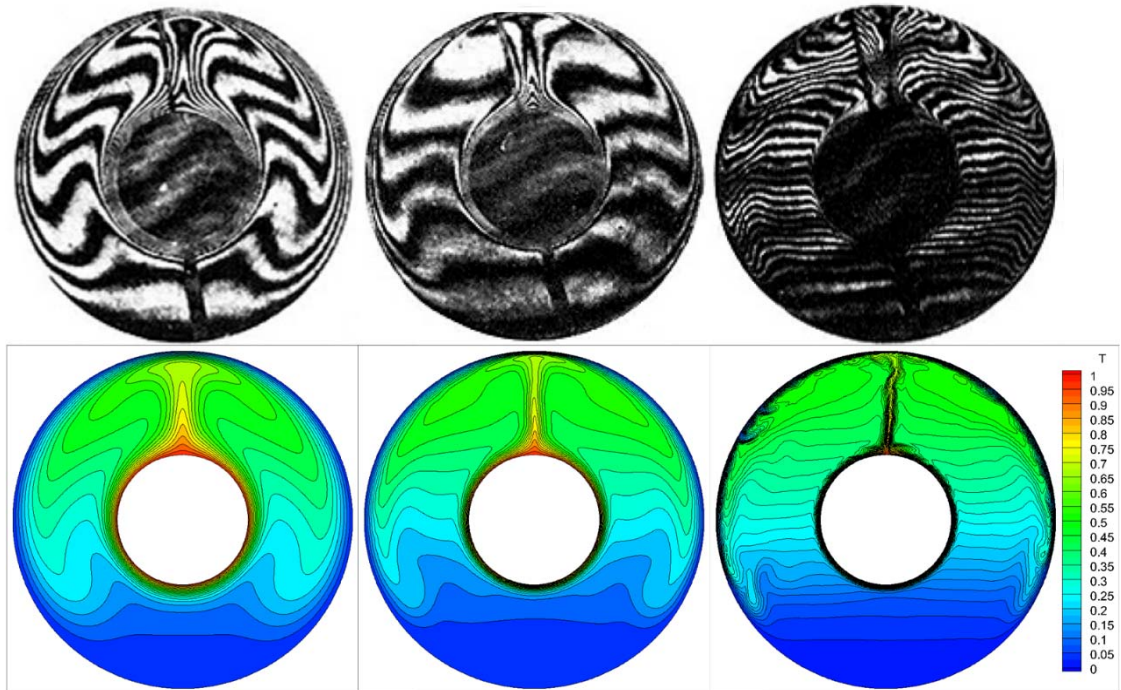

Fig. 4. Comparison of interferograms from experiment (top) [2] and isotherms from SPH simulation (bottom) (left: Pr = 0.718, Ra = 1.02×10$^5$; middle: Pr = 0.723, Ra = 1.06×10$^6$; right: Pr = 0.723, Ra = 1.60×10$^7$).

## 4. Results and discussions

A series of simulations were conducted to study the flow patterns in a horizontal concentric annulus in the range of $10^2 \leq \text{Ra} \leq 10^7$ and $0.01 \leq \text{Pr} \leq 10$. The numerical results of the temperature distributions and streamlines at different Rayleigh and Prandtl numbers are shown in Figs. 5 and 6.

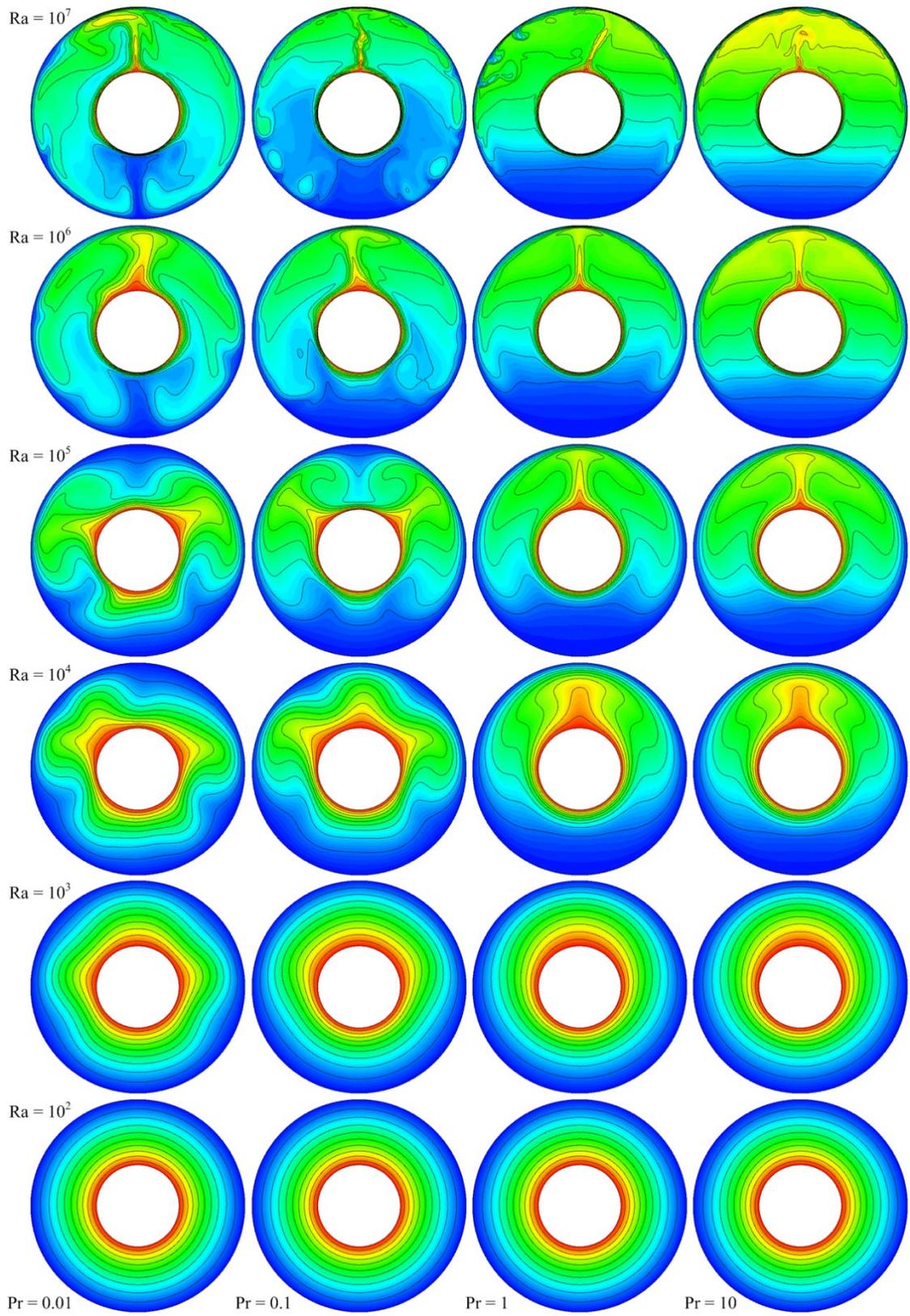

Fig. 5. Temperature distributions at different Prandtl and Rayleigh numbers.

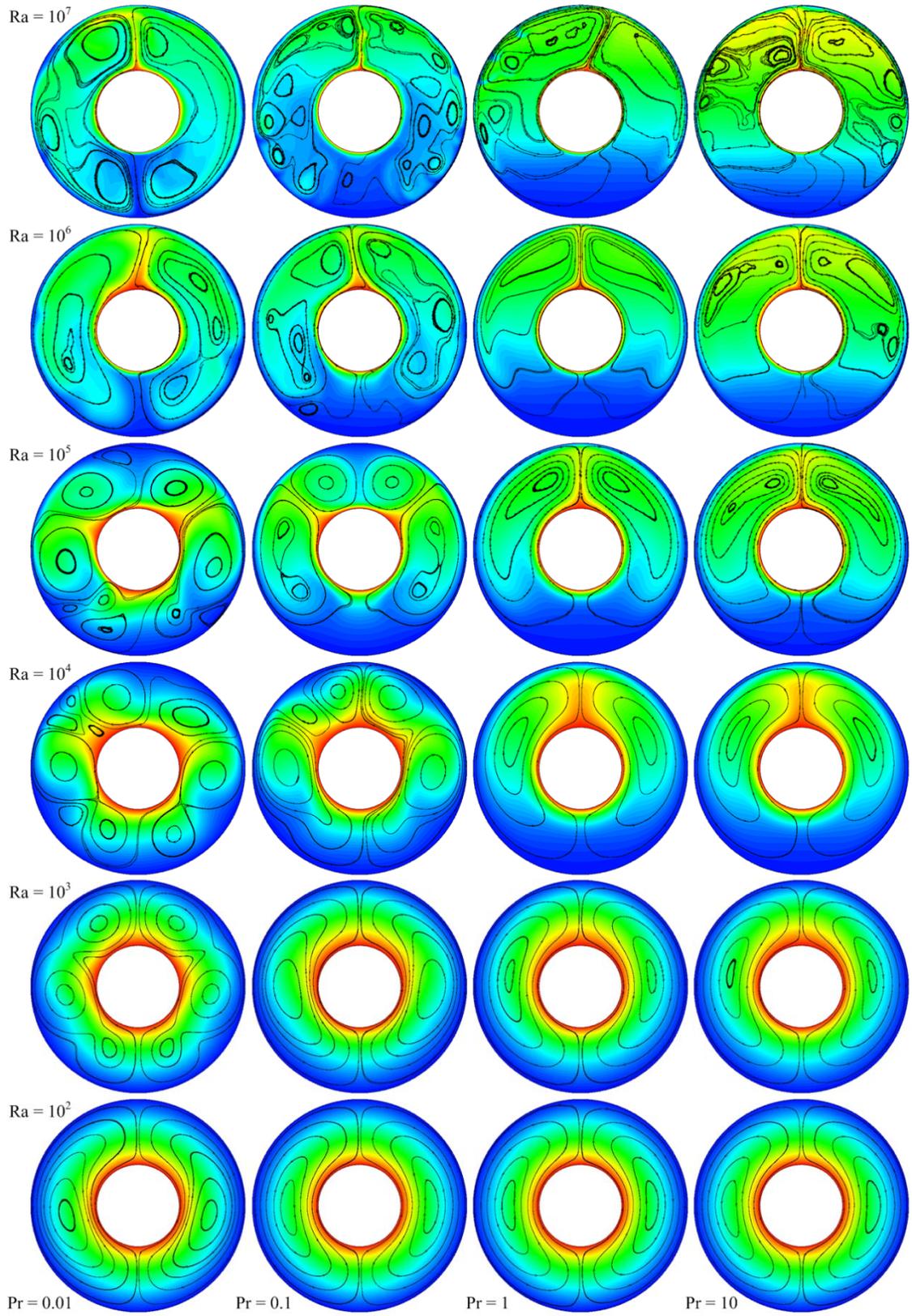

Fig. 6. Streamlines at different Prandtl and Rayleigh numbers.

When the Rayleigh number is $10^2$, the conduction dominates the heat transfer in the annulus and the natural convection is very weak. It can be seen in Fig. 5 that the isotherms are almost concentric circles. Although the natural convection is very weak, the two symmetric vertices caused by natural convection can be seen in Fig. 6. The height of the vortex center increases slightly with the increase of Prandtl number.

At higher Rayleigh number Ra = $10^3$, the natural convection in the annulus becomes stronger, which led to that the isotherms move up and change their shapes. The effect of the Prandtl number on natural convection also becomes more significant. For the cases with Pr = 1 and 10, the isotherms look like eccentric circles, as shown in Fig. 5, while they look like polygons in the cases with Pr = 0.01 and 0.1. Another difference, which is shown in Fig. 6, is that there are six symmetric vertices in the case with Pr = 0.01, while there are only two symmetric vertices in other cases.

When the Rayleigh number is increased to $10^4$, the hot fluid rises and forms plumes in the annulus. Fig. 5 shows that there is only one plume in the cases with Pr = 1 and 10, while there are more plumes in the cases with Pr = 0.1 and 0.01. The plume begins from the hot inner cylinder to the cold outer cylinder, and there is a cold fluid layer between the plume and the outer cylinder. Both the width of the plume and the thickness of the cold layer decrease with the increase of the Rayleigh number from $10^4$ to $10^6$. For the cases with Pr = 0.1 and 0.01, it should be noted that the number of plumes decreases as the Rayleigh number is increased from $10^4$ to $10^6$ and that a plume of cold fluid which begins from the top of the cold outer cylinder to the hot inner cylinder is formed at Ra = $10^5$. This phenomenon is similar to the so-called Rayleigh-Benard convection.

When the Rayleigh number is further increased to $10^7$, the plume is unstable and the flow is turbulent. Figs. 5 and 6 indicate that only the top half part is unstable in the cases with Pr = 1 and 10, while all the areas in the annulus, from top to bottom, is unstable for the cases with Pr = 0.1 and 0.01. The transition Rayleigh number of the fluid from stable to unstable states varies with the Prandtl number.

Based on the numerical results, four convection states can be identified: stable state with 1

plume (SP1), unstable state with 1 plume (UP1), stable state with n (n > 1) plumes (SPN), and unstable state with n (n > 1) plumes (UPN). Fig. 7 shows the convection states as a function of Rayleigh and Prandtl numbers. There are four states at low Prandtl numbers (Pr = 0.1 and 0.01) but only two convection states at high Prandtl numbers (Pr = 1 and 10). The states with more than 1 plumes (SPN and UPN) are not observed in the cases with high Prandtl number. The transition Rayleigh number for low Prandtl number cases from stable state to unstable state is much lower than high Prandtl number cases. In general, the flow usually becomes unstable at high Rayleigh number. However, a higher Rayleigh number does not mean less stable. An exception case is observed that for Pr = 0.1, the flow is unstable at $Ra = 10^4$ but stable at $Ra = 10^5$.

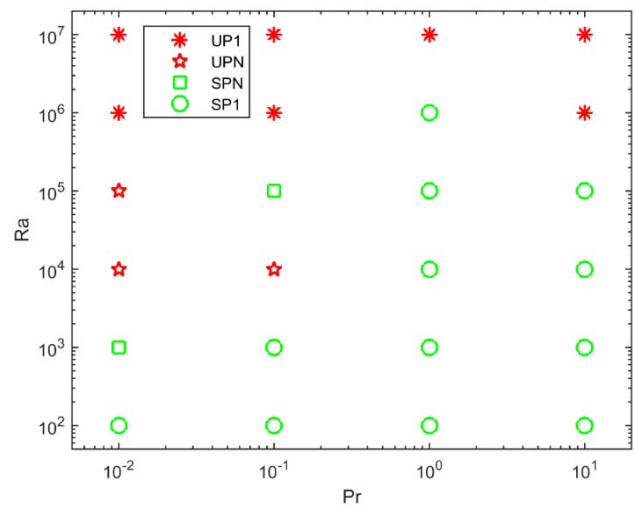

Fig. 7. Convection states with varying Rayleigh and Prandtl numbers. SP1: stable state with 1 plume, SPN: stable state with n (n > 1) plumes, UP1: unstable state with 1 plume, UPN: unstable state with n (n > 1) plumes.

## 5. Conclusions

This paper studied the natural convection states in a horizontal concentric cylindrical annulus using SPH method. A series simulations were conducted in the range of Rayleigh number $10^2 \leq Re \leq 10^7$ and Prandtl number $0.01 \leq Pr \leq 10$. The numerical method was validated by comparing numerical results with experimental results from literature, and good agreement was obtained. Numerical results showed four different convection states: stable state with 1 plume (SP1), unstable state with 1 plume (UP1), stable state with n (n > 1) plumes (SPN), and unstable state with n (n > 1) plumes (UPN). The flow in the annulus is stable at low Rayleigh number but unstable at high Rayleigh number. The transition Rayleigh number from stable to unstable flow varies with Prandtl number. The Prandtl number not only affects the transition Rayleigh number, but also affects the flow patterns. For the cases with low Prandtl numbers, the flow patterns are different and the flow structure are more complex. The SP1 and UP1 states were observed for all Prandtl numbers, while the SPN and UPN states were only observed at Pr = 0.1 and 0.01. The symmetry of the flow is break up at a lower Rayleigh number than the cases with higher Prandtl numbers. The number of vortices at the bottom increases with both Prandtl and Rayleigh numbers. At high Prandtl numbers, the flow instability is mainly at the top area of the annulus. As the Prandtl number decreases, the unstable area increases.

**Acknowledgement**

The authors would like to acknowledge the support of the National Science Foundation (Grant No. CBET-133228).